# Nonextensivity and *q*-distribution of a relativistic gas under an external electromagnetic field

Liu Zhipeng, Du Jiulin and Guo Lina

*Department of Physics, School of Science, Tianjin University, Tianjin 300072, China*

**Abstract**

We investigate the nonextensivity and the *q*-distribution of a relativistic gas under an external electromagnetic field. We derive a formula expression of the nonextensive parameter *q* based on the relativistic generalized Boltzmann equation, the relativistic *q-H* theorem and the relativistic version of *q*-power-law distribution function in the nonextensive *q*-kinetic theory. We thus provide the connection between the parameter $q \neq 1$ and the differentiation of the temperature field of the gas as well as the four-potential with respect to time and space coordinates, and therefore present the nonextensivity for the gas a clearly physical meaning.

## 1. Introduction

In the recent years, Nonextensive Statistical Mechanics (NSM) has been considered as one of the generalizations of Boltzmann-Gibbs (BG) statistics. NSM is studied on the basis of Tsallis entropy [1], dependent on the nonextensive parameter $q$ different from unity, by the form,

$$S_q = k \frac{1-\Sigma_i p_i^q}{1-q}, \qquad (1)$$

where $p_i$ stands for the probability that the system under consideration is in its $i$ th configuration and $k$ is Boltzmann constant. This entropy is nonextensive for the parameter $q \neq 1$ and the deviation of $q$ from unity is said to represent the degree of the nonextensivity. If we take the limit of $q \to 1$, $S_q$ is reduced to the standard BG entropy, $S = k\Sigma_i p_i \ln p_i$, and the extensivity is recovered.

The investigation on NSM has attracted a great deal of attention both from theoretical and observational viewpoints [2-7]. When many traditional theories in BG statistics were continuously generalized in the framework of NSM and were ceaselessly applied to various interesting fields, we needed to know under what circumstances, *e.g.* which class of nonextensive systems and under what physical situation, should NSM be used for the statistical description. So, how to understand the physical meaning of $q$ and how to determine this parameter from the microscopic dynamics of the systems under consideration become the very important problems in NSM and its applications. In these aspects, some theoretical researches have been down for a nonrelativistic gas [8-13], where self-gravitating systems and plasma systems offered the best framework for researching into the nonextensive effects [9-17].

Especially, the formula expression of the nonextensive parameter $q$ has been determined rigorously by using the generalized Maxwell-Boltzmann (MB) distribution, the generalized Boltzmann equation and the *q-H* theorem in the framework of NSM, thus the nonextensivity is related to some nonequilibrium property of the systems with long-range interactions (gravitational forces and Coulombian forces) [10,11,13]. In other words, we have known that, for NSM, the parameter *q* different from unit is related to the temperature gradient and the long-range potentials of the systems such as self-gravitating system and plasma system. Thus NSM can be reasonably applied to describe the thermodynamic properties of the systems under an external field when they are in the nonequilibrium stationary-state. Recently, the above theory has led to an experimental test of NSM by using the solar sound speeds in the helioseismological measurements [17].

Most recently, the generalized Boltzmann equation and the *q-H* theorem for a relativistic case are studied in the *q*-kinetic theory [18]. Along the lines of the kinetic theory of Refs.[10,11], in this letter, we derive the formula expression of the nonextensive parameter $q \neq 1$ for the relativistic case and then present its physical meaning for the relativistic gas under an external electromagnetic field.

## 2. The physical meaning for *q*-parameter

We consider a relativistic gas of *N*-point particles of mass *m* enclosed in a volume *V* and under the action of an external Lorentz four-force field, $F^\mu(x,p) = -\frac{Q}{mc} F^{\mu\nu}(x) p_\nu$, where the particles have four-momentum $p \equiv p^\mu = (E/c, \mathbf{p})$ in each point $x \equiv x^\mu = (ct, \mathbf{r})$ of the space-time, with their energy $E/c = \sqrt{\mathbf{p}^2 + m^2 c^2}$, *Q* is the charge of the particle, $F^{\mu\nu}$ is the Maxwell electromagnetic tensor. The index $\mu$ takes the four values 0,1,2,3, which denotes the time-space coordinates, respectively. The states of this gas can be characterized by a Lorentz invariant one-particle distribution function, $f_q(x,p)$. Thus $f_q(x,p)d^3xd^3p$ gives, at each time $t$, the number of particles in the volume element $d^3xd^3p$ around the space-time position $x$ and momentum **p**. The evolution equation of the relativistic distribution function is assumed to be the relativistic generalized Boltzmann equation in the $q$-kinetic theory [18],

$$p^\mu \partial_\mu f_q + mF^\mu \frac{\partial f_q}{\partial p^\mu} = C_q(f_q), \tag{2}$$

where $\partial_\mu = (c^{-1}\partial t, \nabla)$ is the differentiation of time-space coordinates, respectively, $C_q(f)$ is the relativistic $q$-collisional term. The relativistic version of power-law distribution in the framework of NSM can be obtained as a natural consequence of the relativistic $q$-H theorem,

$$f_q(x,p) = \{1-(1-q)[\alpha(x)+\beta_\mu p^\mu]\}^{1/1-q}, \tag{3}$$

where $\alpha(x)$ and $\beta_\mu(x)$ are arbitrary space and time-dependent parameters. For the above relativistic gas in the presence of an external electromagnetic field, it can be written [18, 19] by

$$f_q(x,p) = nB_q[1-(1-q)(\frac{u-[p^\mu + c^{-1}QA^\mu(x)]U_\mu}{kT})]^{\frac{1}{1-q}}, \tag{4}$$

where $n$ is the number density of the particles, $B_q$ is the normalized constant, $u$ is the Gibbs function per particles, $U_\mu$ is the mean four-velocity of the gas and $A^\mu(x)$ is the four-potential. It is clear that, in the limit $q \to 1$, Eq.(4) is recovered to the well known relativistic Juttner distribution [20],

$$f(x,p) = nB\exp(\frac{u-[p^\mu + c^{-1}QA^\mu(x)]U_\mu}{kT}). \tag{5}$$

To determine the formula express for the nonextensive parameter, according to the relativistic *q-H* theorem, the solution of Eq.(2) will evolve towards the power-law distribution function, Eq.(4), the *q*-collision term vanishes, $C_q(f_q) = 0$, and Eq.(2) is reduced to,

$$p^\mu \partial_\mu f_q + mF^\mu \frac{\partial f_q}{\partial p^\mu} = 0. \tag{6}$$

In other words [21], the distribution function (4) must satisfy Eq.(6). For the sake of convenience, we can write this equation for $f_q^{1-q}(x,p)$ as

$$p^\mu \partial_\mu f_q^{1-q} + mF^\mu \frac{\partial f_q^{1-q}}{\partial p^\mu} = 0. \tag{7}$$

From Eq.(4) we have

$$f^{1-q}(x,p) = n^{1-q} B_q^{1-q} [1-(1-q)(\frac{\mu - [p^\mu + c^{-1}QA^\mu(x)]U_\mu}{k_B T})]. \tag{8}$$

Then,

$$\partial_\mu f^{1-q}(x,p) = [1-(1-q)(\frac{\mu - [p^\mu + c^{-1}QA^\mu]U_\mu}{k_B T})]\partial_\mu (nB_q)^{1-q}$$

$$-[(nB_q)^{1-q}][\frac{-u}{k_B T^2}\partial_\mu T + \frac{p^\mu U_\mu}{k_B T^2}\partial_\mu T + \frac{c^{-1}QA^\mu(x)U_\mu}{k_B T^2}\partial_\mu T - \frac{c^{-1}Q\partial_\mu A^\mu(x)}{k_B T}U_\mu], \tag{9}$$

and

$$\frac{\partial f^{1-q}}{\partial p^\mu} = (nB_q)^{1-q} \frac{U_\mu}{k_B T}. \tag{10}$$

Substitute Eqs.(9) and (10) into Eq.(7), we have

$$p^\mu \partial_\mu (nB_q)^{1-q} - p^\mu (1-q) \frac{u}{k_B T} \partial_\mu (nB_q)^{1-q} + \frac{[p^\mu + c^{-1} Q A^\mu(x)] U_\mu}{k_B T} \partial_\mu (nB_q)^{1-q}$$

$$- p^\mu (nB_q)^{1-q} [\frac{-u + p^\mu U_\mu + c^{-1} Q A^\mu(x) U_\mu}{k_B T^2} \partial_\mu T - \frac{c^{-1} Q \partial_\mu A^\mu(x)}{k_B T} U_\mu]$$

$$- \frac{Q}{c} F^{\mu\nu} p_\nu (nB_q)^{1-q} (1-q) \frac{U_\mu}{k_B T} = 0. \tag{11}$$

We consider that Eq.(11) is the identically null for any arbitrary variable $p$, so the sum of the coefficients of each power for $p$ must be zero. We therefore derive the sum of the coefficients in Eq.(11), for the first-power term of $p$, as

$$p^\mu \partial_\mu (nB_q)^{1-q} - p^\mu (1-q) \frac{u}{k_B T} \partial_\mu (nB_q)^{1-q} + p^\mu (1-q) \frac{c^{-1} Q A^\nu(x) U^\mu}{k_B T} \partial_\mu (nB_q)^{1-q}$$

$$- p^\mu (nB_q)^{1-q} (1-q) [\frac{1}{k_B T^2} (-u + c^{-1} Q A^\mu(x) U_\mu \partial_\mu T - \frac{c^{-1} Q A^\mu(x)}{k_B T} U_\mu] \tag{12}$$

$$- \frac{Q}{c} F^{\mu\nu} p_\nu (nB_q)^{1-q} (1-q) \frac{U_\nu}{k_B T} = 0,$$

and, for the second-power term of $p$, as

$$\partial_\mu (nB_q)^{1-q} - (nB_q)^{1-q} \frac{\partial_\mu T}{T} = 0. \tag{13}$$

Substitute Eq.(13) into Eq.(12), we find the relation

$$\partial_\mu T = (1-q) \frac{Q}{c k_B} (-\partial_\mu A^\nu + F^{\mu\nu}) U_\mu. \tag{15}$$

Furthermore, let $F^{\mu\nu}$ be a dissymmetry tensor and make use of the relation [22],

$$F^{\mu\nu} = \partial_\mu A^\nu - \partial_\nu A^\mu, \tag{16}$$

Eq.(15) becomes

$$\partial_\mu T = (1-q) \frac{-Q}{c k_B} (\partial_\nu A^\mu) U_\mu. \tag{17}$$

Thus we obtain a formula expression of the nonextensive parameter $q \neq 1$ for the relativistic gas under an external electromagnetic field. This formula (17) provides a connection between the parameter $q \neq 1$ and the differentiation of the temperature field of the gas as well as the four-potential with respect to time and space coordinates, and therefore presents the nonextensivity for the gas a clearly physical meaning: we find that $q$ is different from unity if and only if the quantity $\partial_\mu T$ is different from zero. If the temperature field satisfies $\partial_\mu T = 0$, then we have $q = 1$, corresponding to the case of standard BG statistics. If the temperature field is $\partial_\mu T \neq 0$, then we have $q \neq 1$, corresponding to the case of NSM. So, the nonextensive parameter $q \neq 1$ is related closely to the spatial-time inhomogeneity of temperature field of the nonequilibrium relativistic gas under the external electromagnetic field.

## 3. Conclusion

In summary, we investigate the nonextensivity and the power-law distribution of a relativistic gas under an external electromagnetic field. We derive a formula expression of the nonextensive parameter $q$ based on the relativistic generalized Boltzmann equation, the relativistic $q$-H theorem and the relativistic version of $q$-power-law distribution function in the nonextensive $q$-kinetic theory. We thus provide the connection between the parameter $q \neq 1$ and the differentiation of the temperature field of the gas as well as the four-potential with respect to time and space coordinates, and therefore we present the nonextensivity for the gas a clearly physical meaning. The nonextensive parameter $q \neq 1$ is related closely to the spatial-time inhomogeneity of temperature field of the nonequilibrium relativistic gas under the external electromagnetic field.

**Acknowledgements**

This work is supported by the project of the "985" program of TJU of China and also by the National Natural Science Foundation of China under grant 10675088.